\documentclass{article}
\usepackage{amsfonts}
\usepackage{spconf,amsmath,graphicx,epsfig}
\usepackage{subcaption}
\usepackage{mwe}
\usepackage{xcolor}

\title{Characterizing Speech Adversarial Examples Using Self-Attention U-Net Enhancement}
\name{Chao-Han Huck Yang$^{1}$, Jun Qi$^{1}$, Pin-Yu Chen$^{2}$, Xiaoli Ma$^{1}$, Chin-Hui Lee$^{1}$ }
\address{$^1$Electrical and Computer Engineering, Georgia Institute of Technology, Atlanta, GA, USA\\$^2$IBM Research, Yorktown Heights, NY, USA }

\begin{document}
\ninept
\maketitle
\begin{abstract}
Recent studies have highlighted adversarial examples as ubiquitous threats to the deep neural network (DNN) based speech recognition systems. In this work, we present a U-Net based attention model, U-Net$_{At}$, to enhance adversarial speech signals. Specifically, we evaluate the model performance by interpretable speech recognition metrics and discuss the model performance by the augmented adversarial training. Our experiments show that our proposed U-Net$_{At}$ improves the perceptual evaluation of speech quality (PESQ) from 1.13 to 2.78, speech transmission index (STI) from 0.65 to 0.75, short-term objective intelligibility (STOI) from 0.83 to 0.96 on the task of speech enhancement with adversarial speech examples. We conduct experiments on the automatic speech recognition (ASR) task with adversarial audio attacks. We find that (i) temporal features learned by the attention network are capable of enhancing the robustness of DNN based ASR models; (ii) the generalization power of DNN based ASR model could be enhanced by applying adversarial training with an additive adversarial data augmentation. The ASR metric on word-error-rates (WERs) shows that there is an absolute 2.22 $\%$ decrease under gradient-based perturbation, and an absolute 2.03 $\%$ decrease, under evolutionary-optimized perturbation, which suggests that our enhancement models with adversarial training can further secure a resilient ASR system.
\end{abstract}

\section{Introduction}
\label{sec1}
Deep neural networks (DNNs) on many audio and multimedia recognition tasks have attained many state-of-the-art benchmark results~\cite{xu2015regression, xu2013experimental}. However, DNNs have been shown to be vulnerable to small additive noises upon inputs with adversarial examples~\cite{goodfellow2014explaining}. In particular, audio adversarial examples~\cite{carlini2018audio} were proposed based on the gradient~\cite{carlini2018audio} or gradient-free~\cite{khare2019adversarial} optimization on a targeted loss function, which induces DNN models to make incorrect classification (e.g., a malicious text output, as  "open-the-door") or degraded performance.

As a new challenge on  ASR~\cite{carlini2018audio, yakura2018robust}, adversarial examples for ASR are generated with a small maximum-distortion constrained in magnitude (e.g., SNR) that they could be hard to be detected or noticed upon evaluation metrics and listening. Unfortunately, effective model defense and denoising approaches against audio adversarial examples are still under exploration. Yang et al.~\cite{yang2018characterizing} proposed a detection method by using temporal dependency (TD) with a high detection rate (93.6\%) on audio adversarial examples. However, an ASR system with a TD-framework is \textbf{not} capable of correcting adversarial inputs and thus has to abandon many real-time audio examples under a continuous adversarial noise generator. 

Our study is built on DNN-based speech enhancement, which can transform adversarial speech examples into enhanced speech. As shown in \textbf{Fig. 1}, the self-attention enhancement methods are adopted to improve the ASR performance under adversarial perturbation. We summarize the related generative and defense works on audio adversarial examples and speech enhancement as follows:

\textbf{Audio Adversarial Examples.}
To craft adversarial examples against DNN-based ASR systems, security challenges, including speech-to-text and speech classification, have been recently studied by adversary loss optimization. Cisse \emph{et al.}~\cite{cisse2017houdini} introduced a probabilistic loss function to degrade ASR performance. Carlini and Wagner~\cite{carlini2018audio} applied a two-step adversarial optimization with an $l_{\infty}$ norm constrains, which attained a 100\% attacking success rate on ASR. Iter \emph{et al.}~\cite{iter2017generating} leveraged specific extracted audio features like Mel Frequency Cepstral Coefficients (MFCCs) to improve the attacking effectiveness of adversarial examples. Alzantotet \emph{et al.}~\cite{alzantot2018did} and Khare et al. \cite{khare2019adversarial} proposed evolutionary algorithms in a black-box adversarial security setting without accessing the model information of DNN based ASR. However, the previous works did not consider the over-the-air effects in ASR settings as a threat model in the physical world. Qin \emph{et al.} \cite{qin2019imperceptible} proposed an unbounded max-norm optimization to improve the imperceptibility of audio adversarial examples but did not play over-the-air (the experiment works on a room-simulator). 
\begin{figure}[ht!]
\begin{center}
\vspace{-2mm}
   \includegraphics[width=0.90\linewidth]{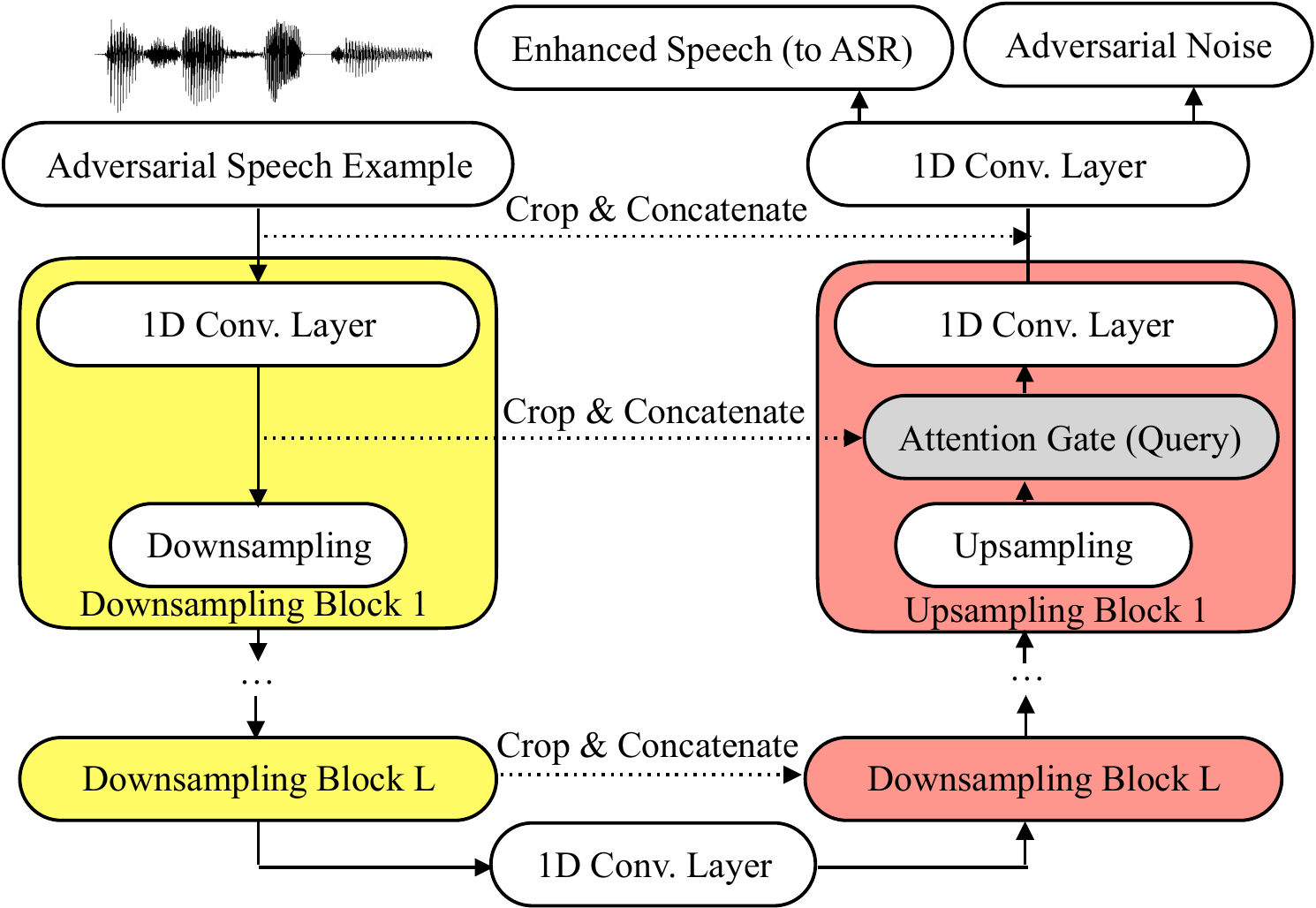}
\end{center}
\vspace{-0.5cm}
   \caption{The proposed a self-attention U-Net (U-Net$_{At}$) structure for improving the adversarial robustness by processing before ASR. %
   } 
\label{fig:figure1}
\end{figure}
\vspace{-2mm}

In recent work, H.Yakura and J. Sakuma~\cite{yakura2018robust} put forth a method to generate a robust adversarial example that could attack DeepSpeech~\cite{hannun2014deep} under the over-the-air condition (e.g., the given adversarial example played by speaker and radio.) In our work, we employ both the gradient and gradient-free adversarial algorithms in our enhancement methods, and try to minimize physical threats of robust adversarial examples in the over the air setting \cite{yakura2018robust}.

\textbf{Speech Enhancement and Denoising Methods.}
 A speech enhancement system aims to improve the speech quality and intelligibility~\cite{xu2013experimental, gravesspeech,  dwang2018}. Several DNN-based vector-to-vector regression architectures~\cite{qi2019theory, xu2015regression} for single-channel speech enhancement have attained many state-of-the-art results under non-stationary noises. However, audio adversarial examples, which are taken as new threats to environments, are not sufficiently discussed whether the deep learning based speech enhancement models can overcome them. Thus, the recent work have started to attempt data augmentation approaches based on adversarial examples to improve the model robustness against adversarial perturbations~\cite{ michelsanti2017conditional, sun2018training}. One major weakness of the adversarial training is that they are highly model-dependent, which means that they may become unstable when there is a strong adversarial perturbation (e.g., an increased magnitude in the gradient-based attacks). In this work, we combine enhancement-based method with temporal features ~\cite{yang2018characterizing}.

To further understand the effects of adversarial examples, we introduce a novel self-attention DNN model, which is sensitive to temporal-sequence segments, and design experiments based on an ASR system to conduct speech enhancement particularly against adversarial perturbations.

\textbf{Our contributions include:}
\begin{itemize}
    \item We introduce speech enhancement-based methods to characterize the properties of audio adversarial examples and improve the robustness of the ASR systems. Specifically, a self-attention U-Net model, U-Net$_{At}$, is introduced to extract the temporal information from the speech with adversarial perturbations, and then to reconstruct speech examples.
    \item We investigate interpretable effects of adversarial examples for speech processing in terms of major speech quality measurement index, including perceptual evaluation of speech quality (PESQ), short-term objective intelligibility (STOI), and speech transmission index (STI). 
    \item We further study the generalization capability of DNN models for exploring the difference of performance among ASR systems via applying both gradient and gradient-free generative models training by adversarial speech data augmentation. 
\end{itemize}
\vspace{-0.3cm}
\section{Audio Adversarial Examples}
\label{sec:sec2}
This section briefly introduces how to use adversarial examples to attack ASR models, which include gradient based~\cite{carlini2018audio}, evolutionary~\cite{khare2019adversarial}, and over-the-air adversarial optimization methods~\cite{yakura2018robust}. 

\subsection{Generating Gradient based Audio Adversarial Examples}
An adversarial example is defined as follows: given a well-trained prediction model $f: \mathbb{R}^{n} \rightarrow\{1,2, \cdots, k\}$ and an input sample $\boldsymbol{x} \in \mathbb{R}^{n}$, an attacker expects to modify $\textbf{x}$ such that the model can recognize the sample as having a specified output $y \in\{1,2, \cdots, k\}$ and the modification does not change the sample significantly. The work \cite{goodfellow2014explaining} proposed $\boldsymbol{v}=\tilde{\boldsymbol{x}}-\boldsymbol{x} \leq \delta $ be the perturbation, where $\delta$ is a parameter with an upper-bounded magnitude of distortion. The distortion is imposed to the input $\textbf{x}$ that humans cannot notice the difference between a legitimate input $\textbf{x}$ and the distorted example $\tilde{\textbf{x}}$. Besides, adversarial examples can be generated by optimizing an objective function as shown in Eq.~(\ref{eq:obj}), where $Loss(\textbf{x}+\textbf{v}, y)$ refers to a loss function by calculating an attacked prediction and the ground truth, and $\epsilon$ is the noise-level:
\begin{equation}
\label{eq:obj}
\underset{\boldsymbol{v}}{\operatorname{argmin}} \operatorname{Loss}(\boldsymbol{x}+\boldsymbol{v}, y)+\epsilon\|\boldsymbol{v}\|
\end{equation}

\textbf{Baseline Audio Adversarial Example.} Particularly, for audio and speech adversarial examples, Mel-Frequency Cepstrum Ceofficient (MFCC)~\cite{carlini2018audio, yakura2018robust,qin2019imperceptible} is used for temporal feature extraction, MFCC can be generated in a gradient-based optimization from an entire waveform using Adam~\cite{kingma2014adam}. In detail, the perturbation $v$ for MFCC can be obtained against the input sample $x$ and the target output of phrase $y$ using the loss function of a ASR system (e.g., DeepSpeech~\cite{hannun2014deep}) as follows:

\begin{equation}
\underset{\boldsymbol{v}}{\operatorname{argmin}} \operatorname{Loss}(M F C C(\boldsymbol{x}+\boldsymbol{v}), \boldsymbol{y})+\epsilon\|\boldsymbol{v}\|
\end{equation}
$MFCC(\textbf{x}+\textbf{v})$ represents the MFCC extraction from speech signals of $\textbf{x}+\textbf{v}$. In the previous work~\cite{carlini2018audio}, this attacking model attains a 100\% success rate on maliciously manipulate the output of speech processing in the non-over-the air condition.

\textbf{Audio Adversarial Examples by Evolutionary Optimization.} Khare \emph{et al.}~\cite{khare2019adversarial} proposed a multi-objective evolutionary optimization method to craft adversarial examples on ASR systems instead of gradient-based approaches~\cite{carlini2018audio, yakura2018robust}. The proposed evolutionary-based method~\cite{khare2019adversarial} focus on maximizing a fitness function, which decomposes the adversarial audio quality metric into two objectives: (a) an Euclidean similarity distance of MFCC features; (b) an edit similarity distance of generated texts. In our experiment, we use the evolutionary optimization combined with the over-the-air loss function in~\cite{yakura2018robust} as a gradient-free comparison (Evo$_{adv}$.)

\textbf{Over-the-Air Adversarial Example.} H.Yakura \& J. Sakuma~\cite{yakura2018robust} proposed a method to generate an over-the-air adversarial example in the real-world. The main modification is to incorporate transformations caused by playback and recording into the generation process, and adapt to three constrains: a band-pass filter, impulse response, and white Gaussian noise to minimize loss function: 

\begin{equation}
\begin{array}{l}{\underset{\boldsymbol{v}}{\operatorname{argmin}} \mathbb{E}_{h \sim \mathcal{H}, \boldsymbol{w} \sim \mathcal{N}\left(0, \sigma^{2}\right)}[\operatorname{Loss}(M F C C(\tilde{\boldsymbol{x}}), \boldsymbol{y})+\epsilon\|\boldsymbol{v}\|]} \\ {\text {where } \overline{\boldsymbol{x}}=\operatorname{Conv}(\boldsymbol{x}+\underset{1000 \sim 4000 \mathrm{Hz}}{\operatorname{BPF}}(\boldsymbol{v}))+\boldsymbol{w}}\end{array}
\end{equation}

where the set of collected impulse responses is $H$ and the convolution using impulse response $h$ is $Conv$, white Gaussian noise is given by $N(0, \sigma^{2})$, and an empirical band-pass filter from 1,000 to 4,000Hz exhibited less distortion. 

In this work, we testify the adversarial generating models like the gradient based \cite{yakura2018robust} (Grad$_{adv}$) and evolutionary generated \cite{khare2019adversarial} (Evo$_{adv}$) examples in the over-the-air filtering setting~\cite{yakura2018robust}. In a more serious adversarial setting, we study the adversarial robustness under adaptive attacks where an ASR model is highly relied on a speech enhancement model, which follows the two-step attack setting in Qin et al.~\cite{qin2019imperceptible}.
\section{Model Defense by U-Net Based Speech Enhancement}
\label{sec:sec3}
U-Net~\cite{ronneberger2015u} refers to deep feature contracting networks by successive layers, where pooling blocks are replaced by up-sampling blocks and large number of feature channels. U-Net was first introduced on image segmentation and attained several state-of-the-art results~\cite{ronneberger2015u}. Recently, Wave-U-Net was proposed by \cite{jansson2017singing} to improve audio source separation and speech enhancement~\cite{macartney2018improved}. However, the previous U-Net-based methods did not consider the sequence-to-sequence mechanism such as temporal dependency. Especially, with the recent evidence~\cite{yang2018characterizing} on the adversarial defense, most audio adversarial examples are with the specific temporal dependency, which indicate that the deep regression-based method~\cite{yang2018characterizing} would remain challenges in correcting or enhancement the adversarial speech examples.   
Our proposed self-attention speech U-Net (U-Net$_{At}$) is a one-dimensional U-Net with down-sampling blocks and a sequential attention gate embedded up-sampling blocks to improve the adversarial robustness as the framwork in \textbf{Fig. 1}.  

\subsection{Self-Attention U-Net for Adversarial Speech Enhancement}
As to a deployment of the U-Net~\cite{jansson2017singing} architecture for speech enhancement, we aim to separate a mixture waveform $m \in[-1,1]^{L \times C}$, as shown in Figure~\ref{fig:figure1}, into source waveform $S_{1}$ and the adversarial waveform $S_{adv}$ $\in[-1,1]^{L \times C}$ for all $k \in {1, 2}$, where C refers to the number of audio channels and $L$ denotes the number of audio samples.  For two sources to be reconstructed, a 1D convolution, zero padded before convolving, of filter size 1 with $2 \times 1$ filters is utilized to convert the stack loss of features of (i) the clean speech example and reconstructed waveform; (ii). We use a block number $L=17, C=1$ for our experiments as the validated results in ~\cite{macartney2018improved}.\\
\textbf{Attention Gate.}
The attention gate in the upsampling block(s) extracts a high-level feature representation $h$ from the input speech feature $x$, where the $h_{t}^{Q}$ and $H_{t}^{K}$ are the query and key, respectively:  
\begin{equation}
\mathbf{h}_{t}^{Q}, \mathbf{H}^{K} =\text { Encoder }(\mathbf{x_{t}});\quad
\mathbf{c}_{t}=\operatorname{Attention}\left( \mathbf{h}_{t}^{Q}, \mathbf{H}^{K},\mathbf{x_c}\right),
\end{equation}
where the attention mechanism takes the query $h_{t}^{Q}$ and key $H_{t}^{k}=[h_{0}, ..., h_{t}]$ as input and devise a fixed-length context vector, $c_{t}$.
The enhanced speech is $\mathbf{y}_{t}$, which takes the context vector $c_{t}$, the upcoming channel input $x_{t}$, and the symmetry downsamping channel block cropping concatenation input $x_{c}$. We adopt the scaled dot-product softmax function for self-attention transformation in ~\cite{vaswani2017attention}. 
\subsection{Speech Enhancement Baseline}

We use the publicly available benchmark LibriSpeech  dataset~\cite{panayotov2015librispeech}. Firstly, we down-sample the speech sampling rate to 16kHz as previous studies~\cite{jansson2017singing}. The clean data consists of $30$ hours training data and $5$ hours testing data, which were from 73 male and female English-speakers with various accents. The noisy data were generated by mixing the clean data with the noise source from the DEMAND noise database~\cite{thiemann2013diverse}. To be consistent with the baseline results in~\cite{macartney2018improved}, we used 40 different noise to corrupt clean speech to generate noisy speech with 4 SNR levels (15, 10, 5, and 0 dB). In sum, there were 8,345 training data and 1,242 test data in our dataset. 
\begin{table}[ht!]
    \centering
\begin{tabular}{ccccc}{\text { \textbf{Metric} }} & {\text { Noisy$_{D}$ }} & {\text { DNN }} & {\text { U-Net$_{W}$ }} & {\text { U-Net$_{At}$ }} \\ \hline \text { PESQ } & {1.97} & {2.62} & {2.86} & {\textbf{2.88}} \\ {\text { STI }} & {0.65} & {0.73} & {\textbf{0.81}} & {\textbf{0.81}}  \\ {\text{STOI}} & {0.82} & {0.90} & \textbf{{0.93}} & {0.92} \\ {\text{SNR}} & {-1.63} & {7.67} & {9.83} & {\textbf{9.85}} \\ \hline\end{tabular}
\vspace{0.1cm}
    \caption{We evaluate the untreated noisy signal (Noist$_{D}$) in ~\cite{thiemann2013diverse}, and the enhanced signals based on DNN, wave U-Net (as U-Net$_{W}$), and self -attention U-Net (as U-Net$_{At}$). The experimental results show that the U-Net based methods attain higher scores compared with DNN-based methods on the noisy speech.}
    \vspace{0.1cm}
    \label{tab:single}
\end{table}
\vspace{-4mm}
\subsection{Performance Analysis}
To validate the general enhancement performance, we evaluate the baseline  performance by major objective indexes as follows:\\
\textbf{SNR.}
We first used the signal-to-noise ratio (SNR) of the perturbation, which could be generated by adding noisy samples~\cite{thiemann2013diverse} or adversarial noise as the Sec. 2. The SNR is given by $10log_{10}P_{x}/P_{v}$ for the power of the input sample, $P_{x}=\frac{1}{T} \sum_{t=1}^{T} x_{t}^{2}$ , and the power of perturbation, $P_{v}=\frac{1}{T} \sum_{t=1}^{T} v_{t}^{2}$ as the previous setting~\cite{yakura2018robust}. \\
\textbf{PSEQ.}
PESQ ~\cite{rix2001perceptual} score is computed as a sum value of the average disturbance $d_{sym}$ and the average  asymmetrical disturbance $d_{asym}$: 
\begin{equation}
P E S Q=a_{0}+a_{1} \cdot d_{s y m}+a_{2} \cdot d_{a s y m}
\end{equation}
where $a_{0}=4.5$, $a_{1}=-0.1$ and $a_{2}=-0.0309$ for interpretability.  
\\
\textbf{STI.}
STI~\cite{steeneken1980physical} is an objective method for prediction and measurement of speech intelligibility, which has been not covered in the previous adversarial studies~\cite{qin2019imperceptible, yakura2018robust, khare2019adversarial} for analyzing speech quality. 
\\
\textbf{STOI.} 
 STOI~\cite{taal2010short} could be used as a robust measurement index for nonlinear processing to noisy speech, e.g., noise reduction on speech intelligibility, but it has been yet evaluated on the adversarial examples.\\
\textbf{Enhancement Effects on Adversarial Examples} After having a well-trained DNN-based SE models, we test the enhancement effects on the generated over-the-air audio adversarial examples \cite{yakura2018robust} against Deep Speech~\cite{hannun2014deep}. Although SNR scores of all models are increased after speech enhancement, the performance in terms of PESQ, STI, and STOI are consistently decreased, which suggests the necessity of conducting a resilient ASR accessibility in \textbf{Table. 1} and \textbf{Table. 2}. 
\begin{table}[ht!]
    \centering
\begin{tabular}{ccccc}{\text { \textbf{Metric} }} & {\text { Noisy$_{adv}$ }} & {\text { DNN }} & {\text { U-Net$_{W}$ }} & {\text { U-Net$_{At}$ }} \\ \hline \text { PESQ } & {\textbf{1.31}} & {1.21} & {1.16} & {1.18} \\ {\text { STI }} & \textbf{{0.67}} & {0.66} & {0.62} & {0.64}  \\ {\text{STOI}} & {\textbf{0.84}} & {0.81} & {0.80} & {0.81} \\ {\text{SNR}} & {-1.52} & {7.23} & {7.43} & {\textbf{7.68}} \\ \hline\end{tabular}
\vspace{0.1cm}
    \caption{We repeat the experiments of speech enhancement in Table 1 where the over-the-air adversarial speech examples (Noisy$_{adv}$) ~\cite{yakura2018robust} were imposed to the ASR loss function. The evaluation results show that although all of the SNR scores increase by the speech enhancement, the other metric indexes are even lower than the Noisy$_{adv}$ before conducting speech enhancement.}
    \vspace{0.1cm}
    \label{tab:tab2}
\end{table}

\textbf{Robust Enhancement by Adversarial Training.}
Goodfellow et al. \cite{goodfellow2014explaining} showed that by training on a mixture of adversarial and clean examples, a neural network could be regularized and robust enough against to adversarial examples. We adopt the adversarial training with an objective function based on the fast gradient sign method as an effective regularizer for a loss function $\tilde{J}$:
\begin{equation}
\tilde{J}(\boldsymbol{\theta}, \boldsymbol{x}, y)=\alpha J(\boldsymbol{\theta}, \boldsymbol{x}, y)+(1-\alpha) J\left(\boldsymbol{\theta}, \boldsymbol{x}+\epsilon \operatorname{sign}\left(\nabla_{\boldsymbol{x}} J(\boldsymbol{\theta}, \boldsymbol{x}, y)\right),\right.
\end{equation}
where $x$ and $y$ separately denote the noisy input and the clean output, and model parameters are represented as $\theta$.
In all of our experiments, we used an empirically fin-tinning setting as $\alpha= 0.34$ to conduct the enhancement results training on the generated adversarial examples. As \textbf{Table. 4}, all the models shows an improved performance with adversarial training. 

\begin{table}[ht!]
    \centering
\begin{tabular}{ccccc}{\text { \textbf{Metric} }} & {\text { Noisy$_{adv}$ }} & {\text {DNN$_{T}$ }} & {\text { U-Net$_{T,W}$ }} & {\text { U-Net$_{T,At}$ }} \\ \hline \text { PESQ } & {1.31} & {2.55} & {2.72} & {\textbf{2.78}} \\ {\text { STI }} & {0.67} & {0.69} & {0.72} & {\textbf{0.75}}  \\ {\text{STOI}} & {0.84} & {0.86} & {0.88} & {\textbf{0.90}} \\ {\text{SNR}} & {-1.52} & {7.45} & {7.67} & {\textbf{7.92}} \\ \hline\end{tabular}
\vspace{0.1cm}
    \caption{To further improve the model generalization capability, we add adversarial examples into the dataset for the adversarial training. We observe that the model obtains a further improvement in terms of the intellectual speech quality based on the methods of the adversarial training on DNN (DNN$_T$), wave U-Net (U-Net$_{T,W}$), and self-attention U-Net (U-Net$_{T,At}$) with slightly improved SNRs.}
    \vspace{0.1cm}
    \label{tab:3}
\end{table}
\section{Experiment Results}
\label{sec:setup}
\subsection{ASR Experiment Setting}
We applied the proposed U-Net$_{At}$ to enhance adversarial speech examples as the same reproducible settings and configurations in \cite{yakura2018robust}. The first clip was the same as the publicly released samples of \cite{carlini2018audio}. For the target phrase $y$, we prepared the target text: \textbf{“open the door”}. Since the previous works \cite{carlini2018audio, yakura2018robust} tested the methods with randomly chosen 1,000 phrases. We follow the over-the-air~\cite{yakura2018robust} evaluation setting with the LibriSpeech dataset~\cite{panayotov2015librispeech} which involves a number of playback cycles in the physical world. 
\subsection{ASR Performance Discussion}
The success rate of the attack is the ratio of the times that Mozilla DeepSpeech~\cite{hannun2014deep}~\footnote{The Mozilla DeepSpeech ASR is the current audio adversarial robustness benchmark~\cite{qin2019imperceptible, carlini2018audio, yakura2018robust} and shows coherent performance including Kaldi ~\cite{khare2019adversarial, qin2019imperceptible} under adversarial attacks from the previous studies. } transcribed the recorded adversarial example as the target phrases among all trials. The success rate becomes non-zero only when Mozilla DeepSpeech transcribes adversarial examples as the target phrases perfectly. The generated audio with high adversarial examples remain a SNR of $10.12$ (dB) for all the experiments. In \textbf{Table. 4}, the model trained on Noisy$_G$ could \textbf{not attain a high WER and RoSA} under Grad$_{adv}$ and Evo$_{adv}$ without the improved performance from the adaptive training with related adversarial examples. \textbf{Fig 2.} shows the spectrogram difference in the recovered detail of speech enhancement and adversarial training.\\

\begin{table}[]
\centering
\begin{tabular}{llllll}
\textbf{Gradient-based~\cite{yakura2018robust}} &w/o $\mathbf{SE}$ & $\mathbf{SE}$~\cite{qi2019theory}  & $\mathbf{SE}$~\cite{macartney2018improved} & $\mathbf{SE}_{U_{At}}$ \\ \hline
\multicolumn{1}{l|}{\textcolor{blue}{RoSA}$_\textbf{DeepSpeech}$} & 90.23 & 84.86  & 83.43 & 83.11 \\
\multicolumn{1}{l|}{RoSA$_\textbf{DeepSpeech+AdvT}$ } & 27.34 & 22.21 & 19.29 & \textbf{18.23} \\ \hline
\multicolumn{1}{l|}{\textcolor{blue}{WER}$_\textbf{DeepSpeech}$} & 85.90 & 72.97 & 67.46 & 66.12 \\
\multicolumn{1}{l|}{WER$_\textbf{DeepSpeech+AdvT}$} & 19.37 & 18.92 & 17.64 & \textbf{17.15} \\ \hline
\textbf{Evolution-based~\cite{khare2019adversarial}} & w/o $\mathbf{SE}$ & $\mathbf{SE}$~\cite{qi2019theory}  & $\mathbf{SE}~$\cite{macartney2018improved} & $\mathbf{SE}_{U_{At}}$ \\ \hline
\multicolumn{1}{l|}{\textcolor{purple}{RoSA}$_\textbf{DeepSpeech}$} & 91.21 & 85.67 & 82.03 & 79.35 \\
\multicolumn{1}{l|}{RoSA$_\textbf{DeepSpeech+AdvT}$} & 20.47 & 18.45 & 17.81 & \textbf{16.14} \\ \hline
\multicolumn{1}{l|}{\textcolor{purple}{WER}$_\textbf{DeepSpeech}$} & 87.90 & 83.12 & 79.20 & 71.12 \\
\multicolumn{1}{l|}{WER$_\textbf{DeepSpeech+AdvT}$} & 19.45 & 19.42 & 18.44 & \textbf{17.42} \\ \hline
\end{tabular}
\caption{To improve the adversarial robustness of the ASR, we utilize adversarial training (\textbf{AdvT}) on the proposed self-attention U-Net ($U_{At}$) speech enhancement ($\mathbf{SE}$). From our experimental results, the use of augmented gradient-based adversarial examples (AEs) can improve the performance in terms of the rate of success attack (RoSA) and the word error rate (WER \%). The evolutionary-optimized~\cite{khare2019adversarial} AEs exhibit severe error rate without \textbf{AdvT}, but WER under \textbf{AdvT} \textbf{can obtain more gains} than the gradient-based adversarial settings~\cite{yakura2018robust}. The deployed adversarial text output is \textbf{``open-the-door''} to evaluate WER based targeted attack scenarios.}
\label{tab:my-table:4}
\end{table}

\label{sec:res}

\begin{figure}[ht!]
        \centering
        \begin{subfigure}[b]{0.230\textwidth}
            \centering
            \includegraphics[width=\textwidth]{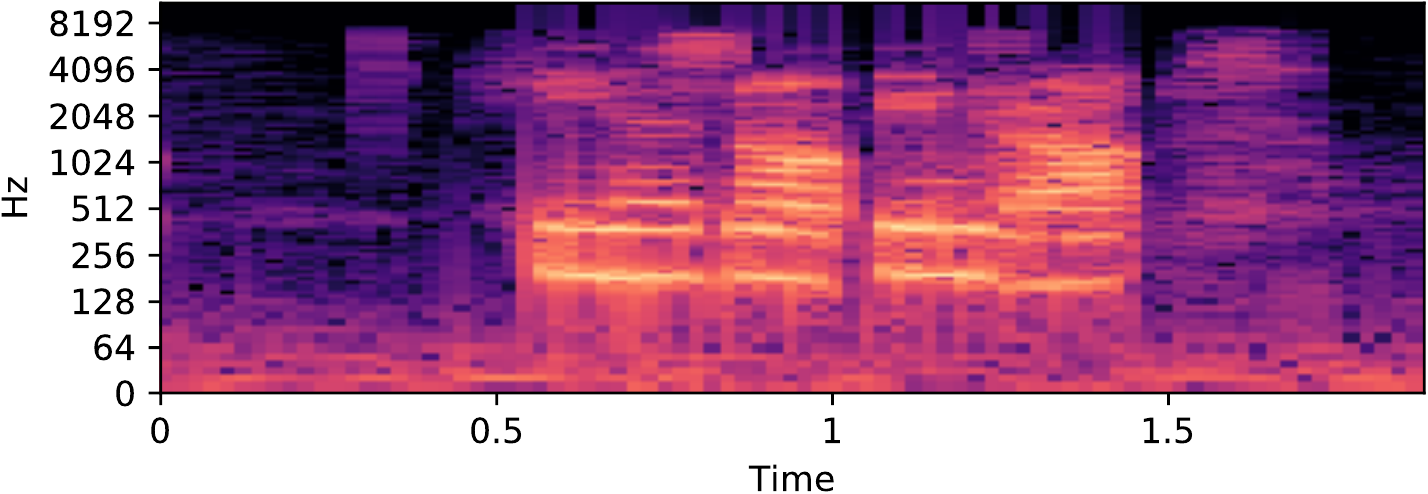}
            \caption[Network2]%
            {{\small Clean}}    
            \label{fig:mean and std of net14}
        \end{subfigure}
        \hfill
        \begin{subfigure}[b]{0.230\textwidth}  
            \centering 
            \includegraphics[width=\textwidth]{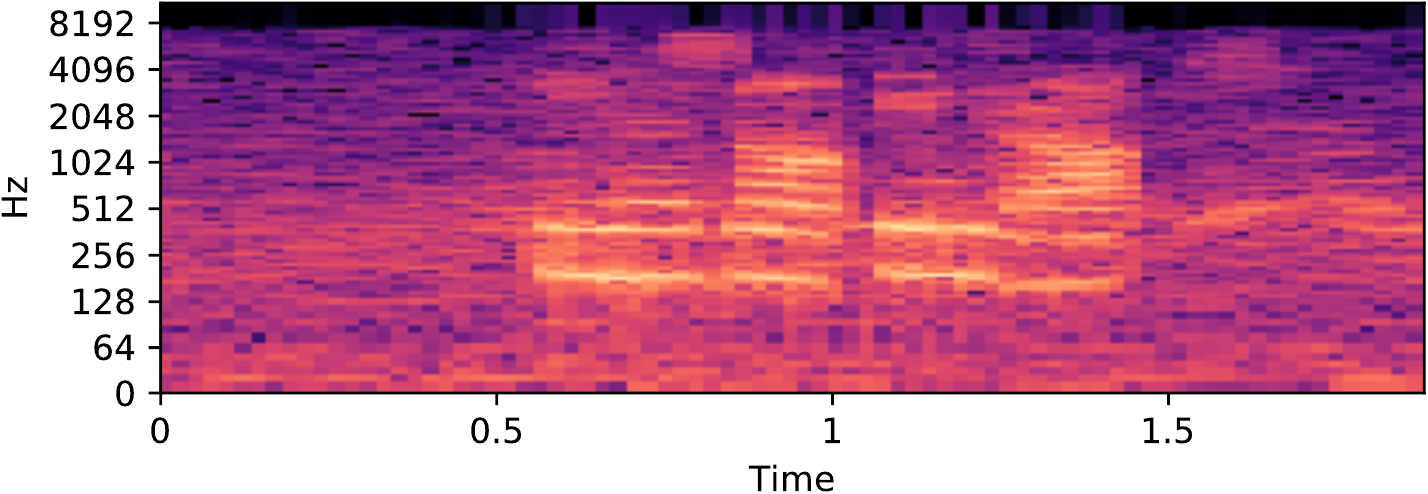}
            \caption[]%
            {{\small Noisy}}    
            \label{fig:mean and std of net24}
        \end{subfigure}
        \vskip\baselineskip
        \begin{subfigure}[b]{0.230\textwidth}   
            \centering 
            \includegraphics[width=\textwidth]{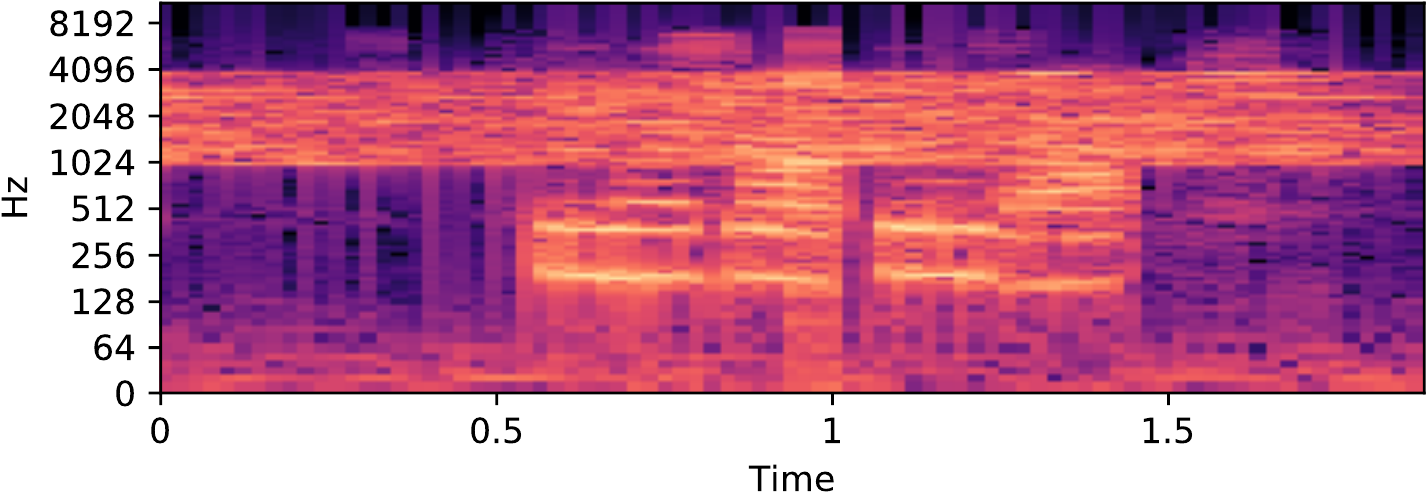}
            \caption[]%
            {{\small Adversarial}}    
            \label{fig:mean and std of net34}
        \end{subfigure}
        \quad
        \begin{subfigure}[b]{0.230\textwidth}   
            \centering 
            \includegraphics[width=\textwidth]{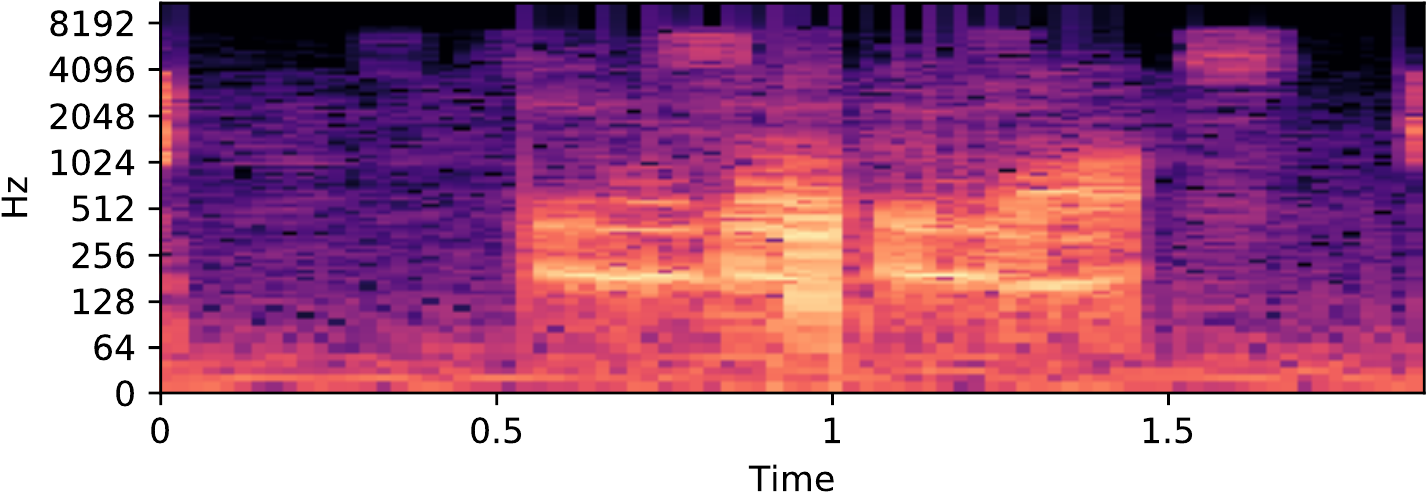}
            \caption[]%
            {{\small U-Net$_{At}$-Enhanced}}    
            \label{fig:mean and std of net44}
        \end{subfigure}
        \vskip\baselineskip
        \begin{subfigure}[b]{0.230\textwidth}   
            \centering 
            \includegraphics[width=\textwidth]{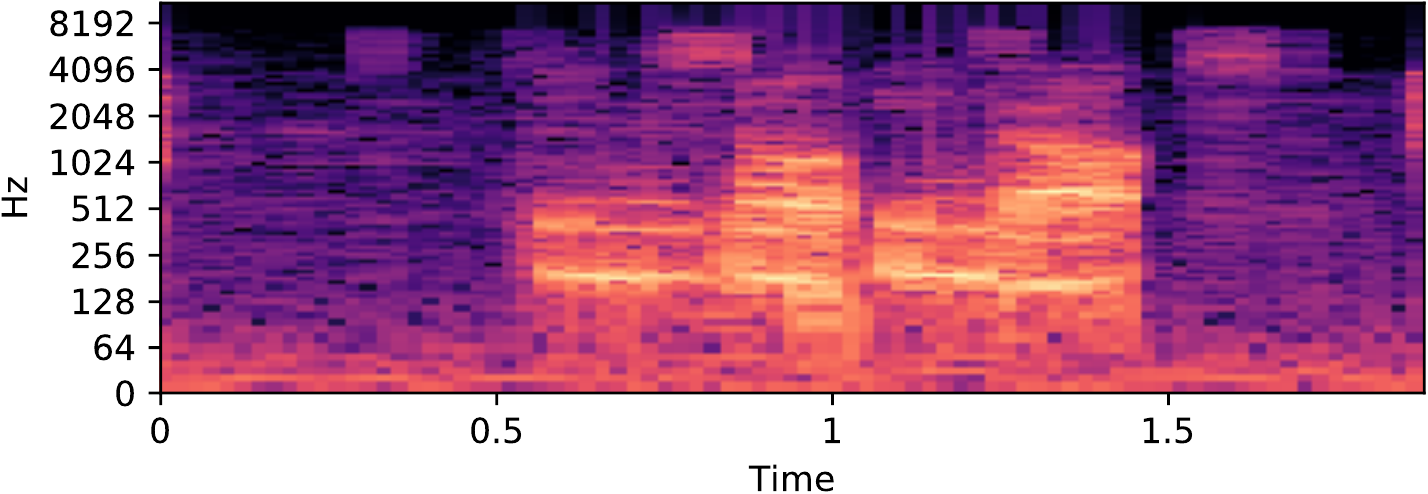}
            \caption[]%
            {{\small AdvT-U-Net$_{W}$-Enhanced}}    
            \label{fig:mean and std of net54}
        \end{subfigure}
        \quad
        \begin{subfigure}[b]{0.230\textwidth}   
            \centering 
            \includegraphics[width=\textwidth]{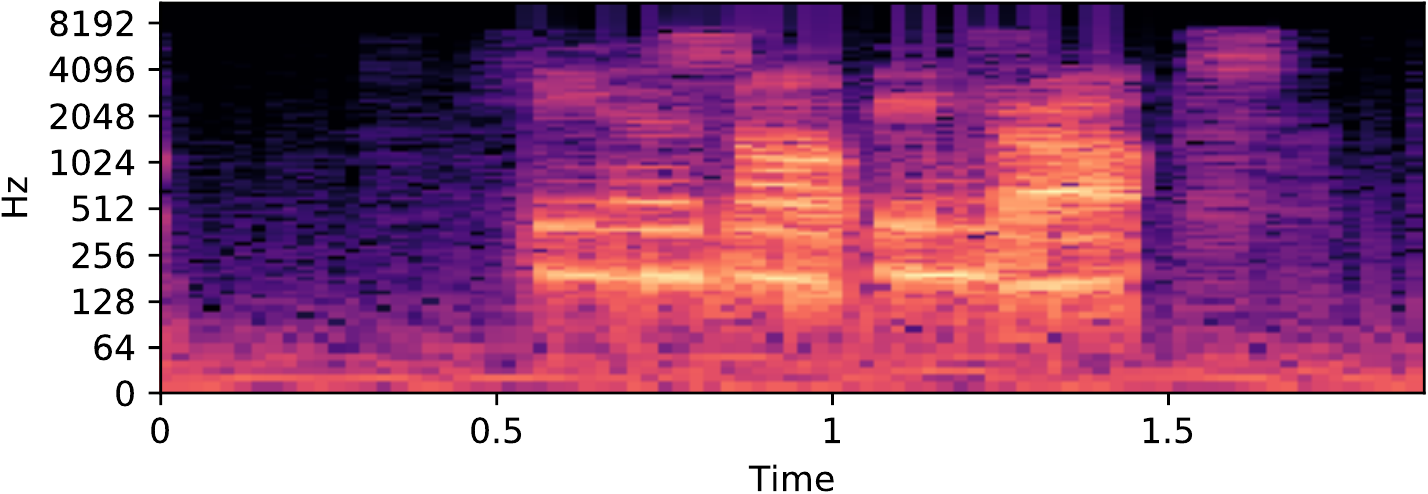}
            \caption[]%
            {{\small AdvT-U-Net$_{At}$-Enhanced}}    
            \label{fig:mean and std of net64}
        \end{subfigure}
        \caption[ The average and standard deviation of critical parameters ]
        {\small The log-power spectrogram of (a) clean; (b) noisy; (c) adversarial; (d) pre-trained U-Net$_{At}$ enhanced adversarial examples; (e) DNN enhancement results adopted adversarial training, and (f) proposed U-Net$_{At}$ using adversarial training (AdvT.)} 
        \label{fig:mean and std of nets}
    \end{figure}

\begin{figure}[ht!]
\begin{center}
\vspace{-2mm}
   \includegraphics[width=0.9\linewidth]{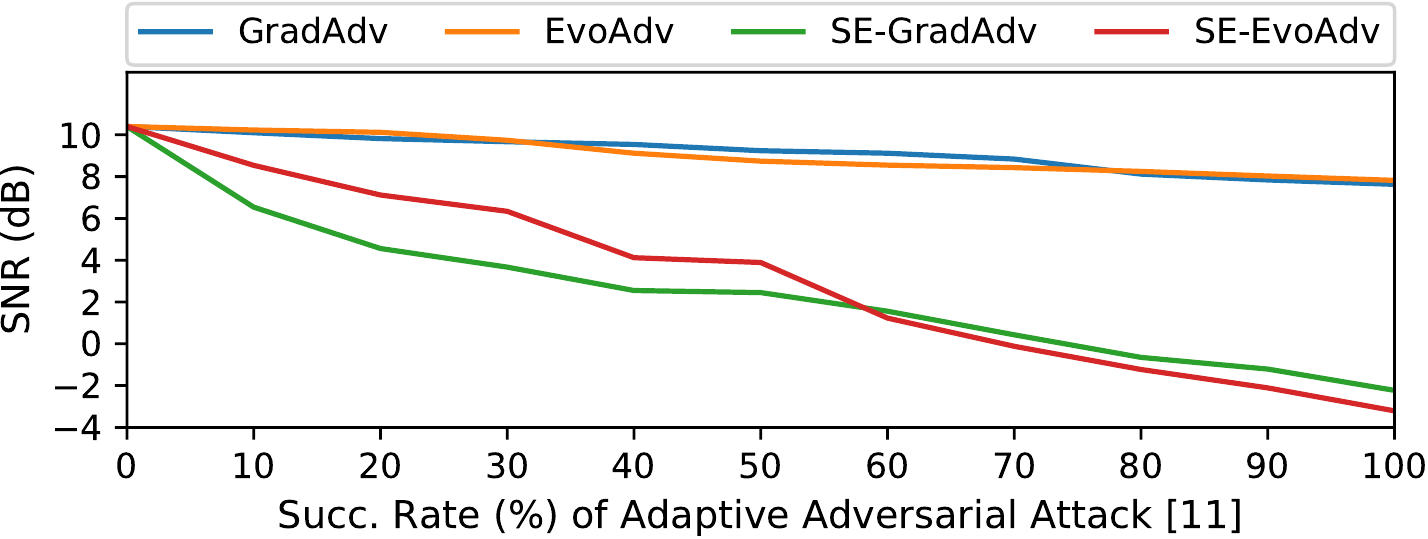}
\end{center}
\vspace{-0.5cm}
   \caption{In a more strict adversarial security setting, we evaluate the SNR magnitude of the Grad$_{Adv}$ and Evo$_{Adv}$ under U-Net$_{At}$ SE. } 
\label{fig:figure3}
\end{figure}
\vspace{-4mm}

\textbf{Characterising the Adaptive Adversarial Robustness.}
In more strict adversarial security setting~\cite{qin2019imperceptible}, an attacker could observe the input $\hat{x}$ and output $\hat{y}$ of both speech enhancement and ASR system\footnote{In our case, the system is equal to pre-processing U-Net$_{At}$ model and Deep Speech ASR for adaptive adversarial attack as ~\cite{qin2019imperceptible}}, to craft a two-step adversarial examples in an existence of defense model. When this theoretical adaptive attack exist, we aim to enlarge the cost of attack in terms of additive noise (dB) injected into the clean speech example to access a targeted attack successful rate (TASR.) In \textbf{Fig. 3}, the results show that an attacker need to give additive noise from 7.63 dB (w/o enhancement) to -2.23 dB (w/ U-Net$_{At}$) for the Grad$_{adv}$ and 7.82 dB (w/o enhancement) -3.21 dB (w/ U-Net$_{At}$) on the Evo$_{adv}$ on the Deep Speech-based ASR system to attain 100\% TASR, which increase the empirical difficulty for attackers to manipulate an attack without notice in the real world and improve the adversarial robustness of ASR system. 
\vspace{-2mm}
\section{Conclusions}
\label{sec:conclusion}
 We demonstrate the power of adversarial speech enhancement by proposed self-attention U-Net, U-Net$_{At}$, for characterizing adversarial examples generated by two state-of-the-art audio adversarial attacks~\cite{yakura2018robust, khare2019adversarial} in the over-the-air setting. The proposed enhancement method is different from previous detection based methods. Our results highlight the importance of ASR and speech processing safety on exploiting unique data properties toward adversarial robustness. Our future works included more acoustic properties analysis.

\clearpage
\bibliographystyle{IEEEbib}
\bibliography{refs}

\begin{thebibliography}{10}

\bibitem{xu2015regression}
Y.~Xu, J.~Du, L.-R. Dai, and C.-H. Lee,
\newblock ``A regression approach to speech enhancement based on deep neural
  networks,''
\newblock {\em IEEE/ACM Transactions on Audio, Speech and Language Processing
  (TASLP)}, vol. 23, no. 1, pp. 7--19, 2015.

\bibitem{xu2013experimental}
Yong Xu, Jun Du, Li-Rong Dai, and Chin-Hui Lee,
\newblock ``An experimental study on speech enhancement based on deep neural
  networks,''
\newblock {\em IEEE Signal processing letters}, vol. 21, no. 1, pp. 65--68,
  2013.

\bibitem{goodfellow2014explaining}
Ian~J Goodfellow, Jonathon Shlens, and Christian Szegedy,
\newblock ``Explaining and harnessing adversarial examples,''
\newblock {\em in International Conference on Learning Representations
  (ICLR),}, 2015.

\bibitem{carlini2018audio}
Nicholas Carlini and David Wagner,
\newblock ``Audio adversarial examples: Targeted attacks on speech-to-text,''
\newblock in {\em 2018 IEEE Security and Privacy Workshops (SPW)}. IEEE, 2018,
  pp. 1--7.

\bibitem{khare2019adversarial}
Shreya Khare, Rahul Aralikatte, and Senthil Mani,
\newblock ``Adversarial black-box attacks on automatic speech recognition
  systems using multi-objective evolutionary optimization,''
\newblock {\em Proc. Interspeech 2019}, pp. 3208--3212, 2019.

\bibitem{yakura2018robust}
Hiromu Yakura and Jun Sakuma,
\newblock ``Robust audio adversarial example for a physical attack,''
\newblock {\em Proceedings of the 28th International Joint Conference on
  Artificial Intelligence, P 5334-5341}, 2019.

\bibitem{yang2018characterizing}
Zhuolin Yang, Bo~Li, Pin-Yu Chen, and Dawn Song,
\newblock ``Characterizing audio adversarial examples using temporal
  dependency,''
\newblock {\em in International Conference on Learning Representations
  (ICLR),}, 2019.

\bibitem{cisse2017houdini}
Moustapha~M Cisse, Yossi Adi, Natalia Neverova, and Joseph Keshet,
\newblock ``Houdini: Fooling deep structured visual and speech recognition
  models with adversarial examples,''
\newblock in {\em Advances in neural information processing systems}, 2017, pp.
  6977--6987.

\bibitem{iter2017generating}
Dan Iter, Jade Huang, and Mike Jermann,
\newblock ``Generating adversarial examples for speech recognition,''
\newblock {\em Stanford Technical Report}, 2017.

\bibitem{alzantot2018did}
Moustafa Alzantot, Bharathan Balaji, and Mani Srivastava,
\newblock ``Did you hear that? adversarial examples against automatic speech
  recognition,''
\newblock {\em arXiv preprint arXiv:1801.00554}, 2018.

\bibitem{qin2019imperceptible}
Yao Qin, Nicholas Carlini, Garrison Cottrell, Ian Goodfellow, and Colin Raffel,
\newblock ``Imperceptible, robust, and targeted adversarial examples for
  automatic speech recognition,''
\newblock in {\em International Conference on Machine Learning}, 2019, pp.
  5231--5240.

\bibitem{hannun2014deep}
Awni Hannun, Carl Case, Jared Casper, Bryan Catanzaro, Greg Diamos, Erich
  Elsen, Ryan Prenger, Sanjeev Satheesh, Shubho Sengupta, Adam Coates, et~al.,
\newblock ``Deep speech: Scaling up end-to-end speech recognition,''
\newblock {\em arXiv preprint arXiv:1412.5567}, 2014.

\bibitem{gravesspeech}
A~Graves, AR~Mohamed, and G~Hinton,
\newblock ``Speech recognition with deep recurrent neural networks, 2013 ieee
  int,''
\newblock in {\em Conf., Acoustics, Speech and Signal Processing (ICASSP)},
  2013, pp. 6645--6649.

\bibitem{dwang2018}
D.~Wang and J.~Chen,
\newblock ``Supervised speech separation based on deep learning: An overview,''
\newblock {\em IEEE/ACM Transactions on Audio, Speech, and Language
  Processing}, vol. 26, no. 10, pp. 1702--1726, 2018.

\bibitem{qi2019theory}
J.~Qi, J.~Du, S.M. Siniscalchi, and C.-H. Lee,
\newblock ``A theory on deep neural network based vector-to-vector regression
  with an illustration of its expressive power in speech enhancement,''
\newblock {\em IEEE/ACM Transactions on Audio, Speech, and Language Processing
  (TASLP)}, vol. 27, no. 12, pp. 1932--1943, 2019.

\bibitem{michelsanti2017conditional}
Daniel Michelsanti and Zheng-Hua Tan,
\newblock ``Conditional generative adversarial networks for speech enhancement
  and noise-robust speaker verification,''
\newblock {\em arXiv preprint arXiv:1709.01703}, 2017.

\bibitem{sun2018training}
Sining Sun, Ching-Feng Yeh, Mari Ostendorf, Mei-Yuh Hwang, and Lei Xie,
\newblock ``Training augmentation with adversarial examples for robust speech
  recognition,''
\newblock {\em arXiv preprint arXiv:1806.02782}, 2018.

\bibitem{kingma2014adam}
Diederik~P Kingma and Jimmy Ba,
\newblock ``Adam: A method for stochastic optimization,''
\newblock {\em arXiv preprint arXiv:1412.6980}, 2014.

\bibitem{ronneberger2015u}
Olaf Ronneberger, Philipp Fischer, and Thomas Brox,
\newblock ``U-net: Convolutional networks for biomedical image segmentation,''
\newblock in {\em International Conference on Medical image computing and
  computer-assisted intervention}. Springer, 2015, pp. 234--241.

\bibitem{jansson2017singing}
Andreas Jansson, Eric Humphrey, Nicola Montecchio, Rachel Bittner, Aparna
  Kumar, and Tillman Weyde,
\newblock ``Singing voice separation with deep u-net convolutional networks,''
\newblock 2017.

\bibitem{macartney2018improved}
Craig Macartney and Tillman Weyde,
\newblock ``Improved speech enhancement with the wave-u-net,''
\newblock {\em arXiv preprint arXiv:1811.11307}, 2018.

\bibitem{vaswani2017attention}
Ashish Vaswani, Noam Shazeer, Niki Parmar, Jakob Uszkoreit, Llion Jones,
  Aidan~N Gomez, {\L}ukasz Kaiser, and Illia Polosukhin,
\newblock ``Attention is all you need,''
\newblock in {\em Advances in neural information processing systems}, 2017, pp.
  5998--6008.

\bibitem{panayotov2015librispeech}
Vassil Panayotov, Guoguo Chen, Daniel Povey, and Sanjeev Khudanpur,
\newblock ``Librispeech: an asr corpus based on public domain audio books,''
\newblock in {\em 2015 IEEE International Conference on Acoustics, Speech and
  Signal Processing (ICASSP)}. IEEE, 2015, pp. 5206--5210.

\bibitem{thiemann2013diverse}
Joachim Thiemann, Nobutaka Ito, and Emmanuel Vincent,
\newblock ``The diverse environments multi-channel acoustic noise database
  (demand): A database of multichannel environmental noise recordings,''
\newblock in {\em Proceedings of Meetings on Acoustics ICA2013}. ASA, 2013,
  vol.~19, p. 035081.

\bibitem{rix2001perceptual}
Antony~W Rix, John~G Beerends, Michael~P Hollier, and Andries~P Hekstra,
\newblock ``Perceptual evaluation of speech quality (pesq)-a new method for
  speech quality assessment of telephone networks and codecs,''
\newblock in {\em 2001 IEEE International Conference on Acoustics, Speech, and
  Signal Processing. Proceedings (Cat. No. 01CH37221)}. IEEE, 2001, vol.~2, pp.
  749--752.

\bibitem{steeneken1980physical}
Herman Jacobus~Marie Steeneken and Tammo Houtgast,
\newblock ``A physical method for measuring speech-transmission quality,''
\newblock {\em The Journal of the Acoustical Society of America}, vol. 67, no.
  1, pp. 318--326, 1980.

\bibitem{taal2010short}
Cees~H Taal, Richard~C Hendriks, Richard Heusdens, and Jesper Jensen,
\newblock ``A short-time objective intelligibility measure for time-frequency
  weighted noisy speech,''
\newblock in {\em 2010 IEEE International Conference on Acoustics, Speech and
  Signal Processing}. IEEE, 2010, pp. 4214--4217.

\bibitem{yang2019causal}
Chao-Han~Huck Yang, Yi-Chieh Liu, Pin-Yu Chen, Xiaoli Ma, and Yi-Chang~James
  Tsai,
\newblock ``When causal intervention meets adversarial examples and image
  masking for deep neural networks,''
\newblock in {\em 2019 IEEE International Conference on Image Processing
  (ICIP)}. IEEE, 2019, pp. 3811--3815.

\end{thebibliography}

\end{document}